\documentclass[11pt,twoside]{article}
\usepackage{moreverb}
\usepackage{epic}
\usepackage{latexsym}

\newtheorem{DefinitioN}{\bf Definition}[section]
\newtheorem{example}{Example}[section]
\newenvironment{definition}[1]
               {\begin{DefinitioN}[#1]~\rm}{\hspace{1cm} \hfill $\Box$\end{DefinitioN}}

\begin{document}
\pagestyle{myheadings}
\markboth{WLPE'01}{An Environment for Non Monotonic Logic Programs}
\title{An Environment for the Exploration of Non Monotonic Logic Programs%
\footnote{In A. Kusalik (ed), Proceedings of the Eleventh Workshop on Logic
Programming Environments (WLPE'01),
December 1, 2001, Paphos, Cyprus. 
COmputer Research Repository (http://www.acm.org/corr/), cs.PL/0111049; 
whole proceedings: cs.PL/0111042.} }
\author{Lu\'\i{}s F. Castro \ \ \ David S. Warren \\
  Computer Science Department \\
  SUNY at Stony Brook \\
  NY 11794 USA \\
  \{\texttt{luis,warren}\}\texttt{@cs.sunysb.edu} \\
  \texttt{http://www.cs.sunysb.edu/~\{luis,warren\}}
}

\maketitle

\begin{abstract}
  
  Stable Model Semantics and Well Founded Semantics have been shown to
  be very useful in several applications of non-monotonic reasoning.
  However, Stable Models presents a high computational complexity,
  whereas Well Founded Semantics is easy to compute and provides an
  approximation of Stable Models.  Efficient engines exist for both
  semantics of logic programs. This work presents a computational
  integration of two of such systems, namely XSB and SMODELS. The
  resulting system is called XNMR, and provides an interactive system
  for the exploration of both semantics. Aspects such as modularity
  can be exploited in order to ease debugging of large knowledge bases
  with the usual Prolog debugging techniques and an interactive
  environment. Besides, the use of a full Prolog system as a front-end
  to a Stable Models engine augments the language usually accepted by
  such systems.
  
\end{abstract}

\section{Introduction}
\label{sec:introduction}

Stable Model Semantics (STABLE) has been shown to be very useful in
several applications in the field of non-monotonic reasoning. However,
computing the stable models semantics is computationally expensive.
In fact, it has been shown that determining whether a given
propositional program has a stable model is already an NP-complete
problem~\cite{Mare88a}.  Well Founded Semantics (WFS) provides an
interesting alternative for non-monotonic applications, presenting an
approximation of stable models with a much lower complexity.
Efficient engines exist that implement both semantics for (classes of)
logic programs. XSB is a Prolog system that relies on a \emph{tabling}
mechanism to implement the Well Founded semantics of normal logic
programs. SMODELS implements Stable Model semantics for a special
class of programs, called \emph{range-restricted} programs, that
impose restrictions on variables and terms appearing in clauses.

In this paper, we present a computational integration of Well Founded
and Stable Model semantics by combining XSB and SMODELS into one
system called XNMR. XNMR aims at providing an interactive environment
for the exploration of non-monotonic semantics of logic programs. The
goal is to provide an easy-to-use environment to study new
possibilities in the joint use of STABLE and WFS.

One important feature of XNMR is its interactive nature. By allowing
the user to interact with the system and experiment with different
queries and semantics, she can obtain a better and deeper
understanding of her program or knowledge base. A related application of
XNMR is in debugging knowledge bases and systems. Additionally
modularity can be exploited for easy experimentation with large
systems. 

The remaining of the paper is divided into three parts. The next
section presents the definitions of both Stable Model semantics and
Well Founded semantics in the common framework of partial Stable
Models. Section~\ref{sec:implementation} presents implementation
issues in XNMR, along with a brief introduction of the characteristics
of the systems it is based on, XSB and SMODELS. Finally,
Section~\ref{sec:environment} presents XNMR from the point-of-view of
the user, presenting its capabilities and features, along with some
caveats resulting from the integration of STABLE and WFS.

\section{Basic Definitions}
\label{sec:semantics}

%Define both STABLE and WFS (probably by means of PSTABLE). Talk about
%the complexities of both semantics, and the non-directiveness of
%STABLE. Discuss using WFS as a pre-processor of STABLE and define the
%class of call-consistent programs as (the/an) interesting one.

The system presented in this paper combines two engines that compute
different semantics for logic programs, namely Well Founded semantics
(WFS) and Stable Model semantics (STABLE). These semantics are
related in that they can both be defined in terms of Partial Stable
Models. In order to provide context for some considerations made in
other parts of the paper we present, in this section, the definitions
for both semantics, in terms of partial stable models.

The \emph{model-theoretic} semantics presented here are defined in
terms of fixed points of a particular operator.

\subsection{Partial Stable Models}
\label{sec:pstable}

Partial Stable Models assign three-valued models to logic
programs. That is, an atom can be assigned truth values of
\emph{true}, \emph{false} or \emph{undefined}. 

Before introducing the definition of Partial Stable Models we show the
\emph{Quotient} operator, first introduced in~\cite{Prz91stable}. The
Quotient operator extends the Gelfond-Lifschitz transformation
(\cite{Gelf88}). Given a logic program $P$ and a partial
interpretation $I_3$, it defines a unique non-negative program $P' =
\frac{P}{I_{3}}$.

\begin{definition}{Quotient Operator}{def:quot}
  Consider a logic program $P$ and a partial interpretation $I_3$ of
  $P$. The \emph{quotient of $P$ modulo $I_3$} is the new program
  $\frac{P}{I_3}$ obtained from $P$ by replacing in every clause of $P$
  all \emph{negative} premises $\neg C$ which are true
  (resp. undefined; resp. false) in $I$ by their corresponding truth
  values \texttt{t} (resp. \texttt{u};
  resp. \texttt{f}). 
\end{definition}

The Quotient operator allows one to obtain a non-negative program from
any given logic program and an interpretation. It is possible to
compute the least partial model of such a program by computing the
least fixed point of the $T_{3P}$ operator. The $T_{3P}$ operator is
an extension of the $T_P$ operator over three-valued interpretations. 

\begin{definition}{Immediate consequence operator
  $T_{3P}$~\cite{Prz91stable}}
  
\label{def:imm}
  Given $P$ a definite logic program, and $I_3$ a partial
  interpretation of $P$. $T_{3P}(I_3) =
  \langle\mathnormal{Pos},\mathnormal{Neg}\rangle$ is the partial
  interpretation where:
  \begin{itemize}
    \item $\mathnormal{Pos} = \{ A \iff \exists (A \leftarrow B_1,...,B_n)
      \in P  \mid   \forall i, B_i \in Pos(I)\}$
    \item $\mathnormal{Neg} = \{ A \iff \forall (A \leftarrow B_1,...,B_n)
      \in P, \exists i \mid  B_i \in Neg(I)\}$
    \end{itemize}
\end{definition}

The $T_{3P}$ operator can be used to obtain the least partial model
(LPM) of a program.

\begin{definition}{Least Partial Model~\mbox{\cite{Prz91stable}}}

  If $P$ is a non-negative program, then the operator $T_{3P}$ has a
  least fixed point which coincides with the least partial model
  $\mathnormal{LPM}(P)$ of $P$, i.e., $\mathnormal{LPM}(P)$ is the
  least partial interpretation (w.r.t. the truth ordering $\preceq$)
  $I_3$ such that $T_{3P}(I_3) = I_3$.
\end{definition}

By combining the three previous operators, we define the $\Phi$
operator that, for a given logic program $P$ assigns to each partial
interpretation $I_3$ the least partial model of the quotient program
$\frac{P}{I_3}$. 

\begin{definition}{$\Phi_P$ operator}
  \label{def:phi}
  Given a program $P$ and a partial interpretation $I_3$ of $P$, we
  define $\Phi_P(I_3)$ as a new partial interpretation given by:
  \begin{displaymath}
    \Phi_P(I_3) = \mathnormal{LPM}(\frac{P}{I_3})
  \end{displaymath}
\end{definition}

Partial stable models are defined as fixed points of the $\Phi$
operator. 

\begin{definition}{Partial Stable Models}
  \label{def:pstable}
  A Partial Stable Model $\mathnormal{PSM}(P)$ of a logic program $P$ is a
  partial interpretation $I_3$ which is a fixed point of the $\Phi$
  operator. 
  \begin{displaymath}
    I_3 = \mathnormal{LPM}(\frac{P}{I_3}) 
  \end{displaymath}
  \begin{displaymath}
    \mathnormal{PSM}(P) = I_3
  \end{displaymath}
    
\end{definition}

\subsection{Well Founded Semantics}
\label{sec:wfs}

The Well Founded Semantics assigns to a logic program $P$ a
three-valued model. The original presentation of WFS (\cite{VGRS:wfs})
provides a constructive definition based on an iterated least fixed
point. However, Przymusinski~\cite{Prz91stable} has shown that WFS can be
equivalently defined in terms of Partial Stable Models. This is the
approach used here.

Two orderings can be defined over partial interpretations. The
\emph{truth ordering} $\preceq$ maximizes the number of true atoms,
whereas the \emph{information ordering} minimizes the amount of
undefined atoms.

\begin{definition}[\cite{PODS89*11}]
  Given two partial interpretations $I_3 =
  \langle\mathnormal{Pos}_I,\mathnormal{Neg}_I\rangle$ and $J_3 =
  \langle\mathnormal{Pos}_J,\mathnormal{Neg}_J\rangle$, we define:
  
  \begin{center}
    $I_3 \preceq J_3 \iff \mathnormal{Pos}_I \subseteq
    \mathnormal{Pos}_J \wedge \mathnormal{Neg}_I \supseteq \mathnormal{Neg}_J$;
  \end{center}
  and
  \begin{center}
    $I_3 \subseteq J_3 \iff \mathnormal{Pos}_I \subseteq
    \mathnormal{Pos}_J \wedge \mathnormal{Neg}_I \subseteq
    \mathnormal{Neg}_J$.
  \end{center}
\end{definition}

The \emph{information ordering} is equivalent to the set-theoretical
\emph{inclusion} ordering. The Well Founded Model of a program is
given by the least Partial Stable Model with respect to the
\emph{information ordering}.

\begin{definition}[Well-Founded Semantics]
  \label{def:wfs}
  The well-founded model of a program $P$ is the $\subseteq$-minimal least
  fixed point of the $\Phi$ operator.

  \begin{displaymath}
    WFS(P) = \mathtt{min}_{\subseteq}\{I \mid \Phi(I) = I\}
  \end{displaymath}
\end{definition}

One important characteristic of Well Founded Semantics is that it
provides a unique model for each program. 

\subsection{Stable Model Semantics}
\label{sec:stable}

The Stable Model Semantics assigns, to a given logic program $P$, zero
or more two-valued models. Programs that contain contradictions have
no models; stratified programs have unique models, and non-stratified
programs may have several models. STABLE is defined as the fixed
points of the $\Phi$ operator which are total, that is, those in which
every atom occurs either in the \texttt{true} set or in the
\texttt{false} set. 

\begin{definition}
  \label{def:stable}
  The Stable Models of a program P are given by a set of the
  interpretations that are fixed points of the $\Phi$ operator and
  in which no atom is undefined.
  \begin{displaymath}
    \mathnormal{STABLE}(P) = \{ I | I = \Phi_P(I) \wedge
    \mathnormal{atoms}(P) \setminus (\mathnormal{Pos}(I)  \cup
    \mathnormal{Neg}(I)) = \emptyset \}
  \end{displaymath}
\end{definition}

An interesting observation is that, if the set of Stable Models
for a given program $P$ is not empty, than all models in the set are
\emph{consistent} with the Well Founded model of $P$. That is, all
atoms that are assigned \texttt{true} or \texttt{false} in WFS have
the same assignments in all stable models; the different stable models
differ only on the assignments to those atoms which are
\texttt{undefined} under WFS.

\section{Implementation of the XNMR System}
\label{sec:implementation}
% Present XSB and SMODELS; give an idea of using XSB to replace SMODELS'
% grounding phase(s). Define residual programs and present the interface
% between XSB and SMODELS in terms of these. Talk about the
% 'skolemization' of variables (?)

In this section we present SMODELS~\cite{smodels:engine}, a Stable
Models computation engine, and XSB~\cite{XSB01}, a Well Founded
Semantics-based Prolog system, along with the issues regarding their
integration. We describe how XSB can be used as a partial evaluator
for programs before stable models are computed. The result of this
partial evaluation, called the \emph{residual program}, is the
\emph{glue} used to combine XSB and SMODELS. The architecture of the
XNMR system, consisting of a two-layered model, is also presented.

\subsection{SMODELS}
\label{sec:smodels}

SMODELS is an implementation of an evaluator for the stable model
semantics for range-restricted programs. It is composed of two
modules, namely a program grounder, and the stable models computation
engine.

The program grounder, called \texttt{lparse}~\cite{smodels:grounder},
computes a grounded version of a range-restricted normal program.  All
variables must have their domains specified by the introduction of
special \emph{domain predicates} in every clause of the program. This
disallows the expression of more general programs, where values of
variables are dependent of the input, thus known only at runtime.
Also, complex structures and lists are not allowed, constraining the
flexibility of the language. One evidence of this constraint is the
impossibility to define a simple predicate like \texttt{append/3},
that appends two lists, creating a new one. The grounder is optimized
so as to not create the whole set of ground instances of the program,
but rather a subset of these that is sufficient to ensure that no
stable models are lost.

The stable models computation engine uses a bottom-up backtracking
search along with a pruning method to implement an efficient algorithm
which has a polynomial space complexity. The pruning method is based
on an approximation technique for stable models which is closely
related to the well-founded semantics~\cite{smodels:engine}.

SMODELS also provides an application program interface (API) that
allows it to be used as a library called from other programs. This API
provides access to the functionalities of the SMODELS evaluation
engine alone, without the grounder capabilities. This functionality is
exploited in the system presented in this paper.

\subsection{XSB}
\label{sec:xsb}

The XSB system is, arguably, the most efficient and well-known
implementation of the well-founded semantics.  XSB implements a fast
WFS computation engine based on SLG resolution~\cite{Chen:1996:TED}.
SLG resolution is performed by succesively applying specific
operations which effectively implement WFS semantics for logic
programs.  

The implementation of XSB is based on the use of the tabulation, or
memoization technique. For each tabled predicate found during the
evaluation process, a table is created where the computed answers for
this call are kept. This ensures that loops are easily detected. The
fixpoint check needed to compute the well-founded semantics is
realized by a \emph{completion} procedure, which detects when all
possible solutions to a predicate have already been computed.

The language accepted by XSB is an extension of full Prolog, thus
allowing the use of the full power of logic variables and unrestricted
structures. Standard predicates like, for instance, \texttt{append/3},
which appends two lists, can be easily programmed by using its usual
definition. Also, the interactiveness of the system allows the user to
work with predicates with infinite models by backtracking through the
answers.

XSB, as a side effect of the well-founded model of a program, produces
a residual program for the given query
\cite{cui99:residual_program,Sago96}. The residual program is a
concretization of the mutual interdependencies among the undefined
atoms in the well-founded model.

\subsection{Residual Program}
\label{sec:resprog}

In XSB, undefined atoms are obtained from \emph{conditional} answers.
An answer is said to be conditional if its set of \emph{delayed
  literals} is not empty, and no operation is applicable. The
concretization of the delayed literals into clauses results in a new
program, called the \emph{residual program}. The residual program is a
representation of the interdependencies among the undefined atoms in
the well founded model of a program, where all true and false literals
in the well-founded semantics are appropriately resolved away.

\begin{figure}[htbp]
  \centering\footnotesize
\begin{boxedverbatim}
nmr| ?- [user].
[Compiling user]
:- table p/1, q/1, r/1, s/1, t/1, u/1.
p(X) :- s(X), tnot(r(X)), tnot(q(X)).
q(X) :- s(X), tnot(p(X)), t(X).
q(X) :- u(X).
s(X) :- t(X).
u(a) :- tnot(u(a)).
t(a).
t(b).
r(c).
end_of_file.
[user compiled, cpu time used: 0.0500 seconds]
[user loaded]
\end{boxedverbatim}
  \caption{Example for residual program computation}
  \label{fig:prog:resid}
\end{figure}

\begin{example}
  \label{ex:residual}
  Consider the program $P$ given in Figure~\ref{fig:prog:resid}.  The
  Well Founded model for $P$ says that the set of positive atoms is
  \{\texttt{s, t}\}, the negative atoms are \{\texttt{r}\} and
  \{\texttt{p, q, u}\} are undefined. Suppose the user asks the query
  \texttt{q(a)}. Figure~\ref{fig:exec:resid} shows how the residual
  program for the query can be obtained by collecting the delayed
  literals for the query and recursively for the delayed literals
  themselves.

  \begin{figure}[htbp]
    \centering\footnotesize
\begin{boxedverbatim}
nmr| ?- q(a).

DELAY LIST = [tnot(p(a))]
DELAY LIST = [u(a)] ? ;

no
nmr| ?- p(a).

DELAY LIST = [tnot(q(a))] ? ;

no
nmr| ?- u(a).

DELAY LIST = [tnot(u(a))] ? 

yes
nmr| ?- 

\end{boxedverbatim}
      \caption{Inspecting the residual program}
      \label{fig:exec:resid}
    \end{figure}
%XNMR computes the residual program as shown in
%  Figure~\ref{fig:exec:resid}. First, it collects the delayed goals
%  for the query, as shown in the first result, creating new clauses
%  for the query goal. Then, it recurses over
%  the new clauses collecting their delayed goals and creating their
%  corresponding new clauses. The resulting residual program is shown
%  in Figure~\ref{fig:final:resid}. Notice that this is actually done
%  transparently to the user, when he asks for any interaction with
%  stable models.
    
    The residual program, shown in Figure~\ref{fig:res_prg} is
    constructed by creating clauses associating each literal with its
    set of delayed literals.

  \end{example}

Example~\ref{ex:residual} shows what kind of information a residual
program carries. 

\begin{figure}[htbp]
  \centering\footnotesize
\begin{boxedverbatim}
q(a) :- tnot(p(a)).
q(a) :- u(a).
p(a) :- tnot(q(a)).
u(a) :- tnot(u(a)).
\end{boxedverbatim}
  \caption{Residual Program}
  \label{fig:res_prg}
\end{figure}

\subsection{XNMR}
\label{sec:xnmr}

In Section~\ref{sec:stable} we noted that, when the set of Stable
Models for a program is not empty, they differ only in those atoms
which are undefined under WFS. A residual program represents the
relationship amongst the undefined atoms in a program under WFS. So it
seems natural to use the residual program of a given logic program in
order to compute its stable models. That is exactly the feature
exploited in XNMR.

XNMR provides an interactive system for the exploration of both Well
Founded and Stable Model semantics of logic programs. It uses XSB to
implement a front-end interface for SMODELS. XSB replaces the grounder
of SMODELS, augmenting it with an interactive top-level system similar
to normal Prolog environments. Also, XSB provides a strictly more
powerful grounder\footnote{Provided that literals are ordered so as to
  be consistent with the left-to-right evaluation strategy of Prolog.}
for SMODELS, since it allows the use of complex
structures in the program like, for example, lists. These structures
are grounded \emph{on demand} according to the query given. In most
cases, results for queries in the well-founded semantics are already
ground. In those cases in which that is not true, variables are
\emph{skolemized} by XNMR, which is a semantically sound operation.

One drawback is related to the use of variables in negative literals.
In XSB, a negative goal with an unbound variable \emph{flounders},
whereas in SMODELS, it is easy to deal with this problem since
variables have their domains specified. The system also provides an
application program interface (API) that can be used by Prolog
programs to interact with SMODELS.

XNMR is implemented by exploiting XSB's foreign predicate
interface~\cite{castro99:fpi} and SMODELS' API. These are connected by
a small amount of C \emph{glue code} layer that interfaces both
systems. The basic functionality is complemented by a collection of
Prolog predicates that automate the collection of the residual program
from the internal information in XSB and convert it into a
representation required by SMODELS. The user interface is implemented
by a top-level shell very similar to those of normal Prolog systems,
but with extended capabilities.

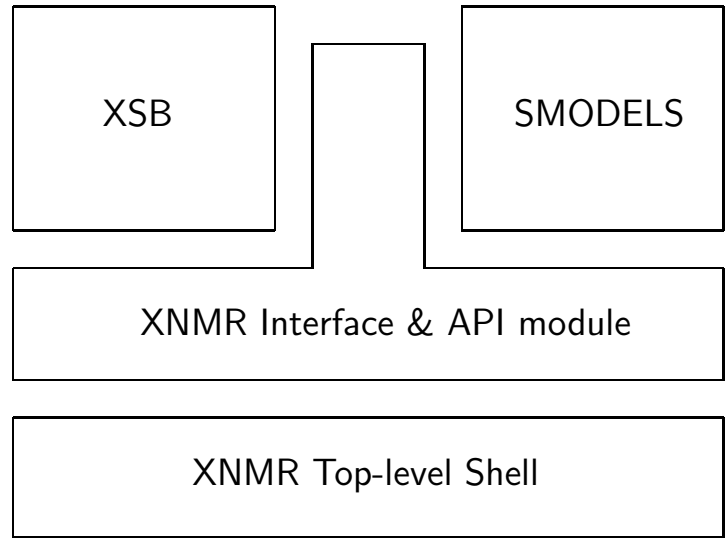
\begin{figure}[htbp]
  \centering\footnotesize
\setlength{\unitlength}{0.00087489in}
\begingroup\makeatletter\ifx\SetFigFont\undefined
% extract first six characters in \fmtname
\def\x#1#2#3#4#5#6#7\relax{\def\x{#1#2#3#4#5#6}}%
\expandafter\x\fmtname xxxxxx\relax \def\y{splain}%
\ifx\x\y   % LaTeX or SliTeX?
\gdef\SetFigFont#1#2#3{%
  \ifnum #1<17\tiny\else \ifnum #1<20\small\else
  \ifnum #1<24\normalsize\else \ifnum #1<29\large\else
  \ifnum #1<34\Large\else \ifnum #1<41\LARGE\else
     \huge\fi\fi\fi\fi\fi\fi
  \csname #3\endcsname}%
\else
\gdef\SetFigFont#1#2#3{\begingroup
  \count@#1\relax \ifnum 25<\count@\count@25\fi
  \def\x{\endgroup\@setsize\SetFigFont{#2pt}}%
  \expandafter\x
    \csname \romannumeral\the\count@ pt\expandafter\endcsname
    \csname @\romannumeral\the\count@ pt\endcsname
  \csname #3\endcsname}%
\fi
\fi\endgroup
{\renewcommand{\dashlinestretch}{30}
\begin{picture}(4299,3234)(0,-10)
\drawline(12,3207)(1587,3207)(1587,1857)
        (12,1857)(12,3207)
\drawline(2712,3207)(4287,3207)(4287,1857)
        (2712,1857)(2712,3207)
\drawline(12,1632)(1812,1632)(1812,2982)
        (2487,2982)(2487,1632)(4287,1632)
        (4287,957)(12,957)(12,1632)
\drawline(12,732)(4287,732)(4287,12)
        (12,12)(12,732)
\put(552,2487){\makebox(0,0)[lb]{\smash{{{\SetFigFont{14}{16.8}{sf}XSB}}}}}
\put(3027,2487){\makebox(0,0)[lb]{\smash{{{\SetFigFont{14}{16.8}{sf}SMODELS}}}}}
\put(777,1227){\makebox(0,0)[lb]{\smash{{{\SetFigFont{14}{16.8}{sf}XNMR Interface \& API module}}}}}
\put(1092,327){\makebox(0,0)[lb]{\smash{{{\SetFigFont{14}{16.8}{sf}XNMR Top-level Shell}}}}}
\end{picture}
}
  \caption{XNMR Architecture}
  \label{fig:arch}
\end{figure}

Figure~\ref{fig:arch} shows the conceptual architecture of the XNMR
system. Both XSB and SMODELS are integrated by the \emph{Integration
  and API module} which also provides a basic API to higher level
layers. The \emph{Top-level shell} is built upon this API and provides
the main interface to the user, by means of an extended Prolog-like
shell.

During the conversion phase, the residual program is collected from
the conditional answers in the tables, and each atom is grounded and
converted to a numerical value, as required by SMODELS. The grounding
is necessary because conditional answers may contain variables. These
variables are \emph{skolemized} by binding them to new constants not
appearing in the program otherwise. After this converted program is
passed to SMODELS, and its stable models are received back, the
conversion phase translates the results back in terms of the original
program, so that they can be displayed to the user.

\section{Functionalities of the XNMR System}
\label{sec:environment}

% Present XNMR as a two-level architecture; the top-level provides the
% interactive environment, and the library as a programming
% interface. Describe functionalities of the top-level (presentation of
% residual programs w/ answers, and optional refinement of answers via
% SMODELS). Describe the library interface.

As shown in the previous section, XNMR combines XSB and SMODELS using
a two-layered architecture. In this section, we describe these layers
from the point of view of the user. We focus mainly in the Top-level
shell, which aims at providing an interface for the end-user. We also
briefly discuss the programmatic interface provided by XNMR's API.

% The XNMR Top-level shell module replaces the standard interactive mode
% of XSB with a specialized top-loop. The XNMR top-loop works much like
% the Prolog standard top-loop, adding some extended functionalities.
% Among these extra capabilities, it provides ways for computing
% different semantics of residual programs derived for undefined
% queries.

\subsection{The Interactive Top-Level Shell}
\label{sec:shell}

The XNMR top-level shell module replaces the standard interactive mode
of XSB with a specialized shell. The XNMR shell mimics standard Prolog
environments, adding features to those common in such systems. These
features include the displaying of delayed literals in undefined
answers to queries, and the possibility of incremental exploration of
stable models for the resulting residual programs.

The prompt works much like a standard Prolog top-level. The difference
is that, when the user issues a query whose value under WFS is
undefined, the system shows the \emph{delay list} for the query. In
this case, the user can either ignore the delay list and assume the
query (with the given instantiations for the variables) is undefined
under the well-founded semantics, or it can ask XNMR for further
information on the query.

Three options are available, which are bound to different keys:

\begin{enumerate}
  
\item[`s'] computes and displays all stable models of the residual
  program, one at a time, by backtracking through them;

\item[`t'] computes and displays those stable models of the residual
  program where the query is true, also backtracking through them, if
  more than one exists; and
  
\item[`a'] checks whether the query is true in all stable models of
  the residual program.

\end{enumerate}

\begin{figure}[htbp]
  \centering\footnotesize
%  \scriptsize
  \begin{boxedverbatim}
$ xsb xnmr
[xsb_configuration loaded]
[sysinitrc loaded]
[packaging loaded]
[sModels loaded]

XSB Version 2.4 (Bavaria) of July 13, 2001
[i686-pc-linux-gnu; mode: optimal; engine: chat; gc: copying; scheduling: local]
nmr| ?- [user].
[Compiling user]
:- table p/0, q/0.
p :- tnot(q).
q :- tnot(p).
end_of_file.
[user compiled, cpu time used: 0.2210 seconds]
[user loaded]

yes
nmr| ?- p.

DELAY LIST = [tnot(q)] ? s

Stable Models: 
  {q} ? ;

  {p} ? ;
  no

no
nmr| ?- p.

DELAY LIST = [tnot(q)] ? t

Stable Models: 
  {p} ? ;
  no

no
nmr| ?- p.

DELAY LIST = [tnot(q)] ? a
  no

no
nmr| ?- halt.
  \end{boxedverbatim}
%$
  \caption{Simple session of XNMR}
  \label{fig:xnmr1}
\end{figure}

Figure~\ref{fig:xnmr1} shows a simple session of the XNMR shell. A
simple program which consists of two atoms that depend negatively on
each other is consulted, and the truth value of one of these atoms is
queried. First, XNMR responds with a \emph{delay list}, which means
that the atom is undefined under WFS. The user is then prompted for a
subsequent action, and the three possibilities are shown. The first
returns the two stable models of the relevant program, one at a time.
The second option displays only that model in which the atom queried
is true. And, finally, the system says \emph{no} for the third option,
meaning that the atom is not true in all stable models of the relevant
program.

\begin{figure}[htbp]
  \centering\footnotesize
%  \scriptsize
  \begin{boxedverbatim}
XSB Version 2.4 (Bavaria) of July 13, 2001
[i686-pc-linux-gnu; mode: optimal; engine: chat; gc: copying; scheduling: local]
nmr| ?- [user].
[Compiling user]
:- table p/0, q/0, r/0.
p :- tnot(q).
q :- tnot(p).
r :- p, tnot(r).
end_of_file.
[user compiled, cpu time used: 0.2590 seconds]
[user loaded]

yes
nmr| ?- p.

DELAY LIST = [tnot(q)] ? s

Stable Models: 
  {q} ? ;

  {p} ? ;
  no

no
nmr| ?- halt.

End XSB (cputime 0.47 secs, elapsetime 54.24 secs)
  \end{boxedverbatim}
  \caption{Relevancy of XNMR results}
  \label{fig:xnmr2}
\end{figure}

Figure~\ref{fig:xnmr2} shows an example of a possibly unexpected
result of the XNMR system. The shown program has only one model under
STABLE, namely the one in which \texttt{q} is true, and \texttt{p} and
\texttt{r} are false. But by computing the residual program of the
query \texttt{p}, only the relevant part of the program is considered,
so two stable models are returned.  It is important to note that XNMR
computes partial stable models of the residual program relevant to a
given query. That means that, as shown above, it is not guaranteed
that every model computed by XNMR corresponds to a total stable model
of the original program. For instance, the program may contain
\emph{inconsistencies} that force it to have no stable models. Still,
if these inconsistencies are not relevant to a given query, XNMR may
return stable models for the residual program that correspond to
partial stable models of the whole program in which the atoms involved
in the inconsistencies are deemed undefined. Another trivial example
is provided in Figure~\ref{fig:moremodels}. The program has only one
stable model, namely the one shown for the query \texttt{r}. That is
due to the inconsistency created by the third clause if we take
\texttt{p} to be true. This effectively forces \texttt{p} to be false
in all stable models. But if we ask the query \texttt{p} to XNMR, as
shown in Figure~\ref{fig:xnmr2} for the same program, the third clause
is not considered when collecting the residual program, thus two
models are computed.

\begin{figure}[htbp]
  \centering\footnotesize
\begin{boxedverbatim}
nmr| ?- [user].
[Compiling user]
:- table p/0, q/0, r/0.
p :- tnot(q).
q :- tnot(p).
r :- p, tnot(r).
end_of_file.
[user compiled, cpu time used: 0.0610 seconds]
[user loaded]

yes
nmr| ?- r.

DELAY LIST = [p,tnot(r)] ? s

Stable Models: 
  {q} ? ;
  no

no
nmr| ?- 
\end{boxedverbatim}
  \caption{Non-relevant inconsistency}
  \label{fig:moremodels}
\end{figure}

This issue of non-relevancy in Stable Model Semantics is well known,
and partly responsible for the non-existence of goal-directed stable
model computation engines. A class of programs has been devised that
is guaranteed to not suffer from these non-relevant inconsistencies.
Programs in this class are called \emph{call-consistent
  programs}~\cite{kenneth89signed,sato90completed}. All models
computed by XNMR for queries over such programs correspond to total
stable models of the original program.

Even though XNMR may compute more models than there are stable models
for a given program, it has been shown in \cite{castro00:rpe} that all
stable models of the program are always represented among the answers
computed. So, if an expected answer is not computed by XNMR, the user
is guaranteed that is also does not appear in any stable model of the
original program. This is an important feature that can be applied in
the debugging of large knowledge bases, specially when combined with
the modularity that is usually necessary to maintain such large
systems.  Individual modules can be inspected separately, forcing
external predicates to be undefined, for instance.

\begin{figure}[htbp]
  \centering\footnotesize
\begin{boxedverbatim}
:- table win/2, move/3.
win(X,L) :- move(X,Y,L), tnot(win(Y,L)).
move(X,Y,L) :- member(m(X,Y),L).
\end{boxedverbatim}
  \caption{Win -- not win program}
  \label{fig:win-notwin}
\end{figure}

The XNMR top-level allows the use of full Prolog in conjunction with
Stable Models computation. As a result of that, complex structures can
be used that will be instantiated and grounded during WFS computation,
before being passed to SMODELS.  The integration of STABLE computation
in the query-based Prolog top-level allows the inspection of programs
that generate possibly infinite structures (or lists) by backtracking,
for instance, interactively until the user feels confident that the
program is computing the desired result. Additionally, the
interactivity of the query-based shell, with unrestricted variables,
allows for much more flexible programs. For instance, consider the
classic \texttt{win - not win} program over a graph. In XNMR, it can
be modeled to have the graph of possible moves encoded as a list, as
shown in Figure~\ref{fig:win-notwin}. The size and shape of the graph
can be changed by asking different queries, without any modifications
to the program. Figure~\ref{fig:win-notwin-rt} shows some possible
queries.

\begin{figure}[htbp]
  \centering\footnotesize
\begin{boxedverbatim}
nmr| ?- win(a,[m(a,b), m(b,c), m(c,d)]). % odd chain

yes
nmr| ?- win(a,[m(a,b), m(b,c), m(c,d), m(d,e)]). % even chain

no
nmr| ?- win(a,[m(a,b), m(b,c), m(c,d), m(d,e), m(e,a)]). % odd cycle

DELAY LIST = [tnot(win(b,[m(a,b),m(b,c),m(c,d),m(d,e),m(e,a)]))] ? s

Stable Models:   no

no
nmr| ?- win(a,[m(a,b),m(b,c),m(c,d),m(d,a)]). % even cycle

DELAY LIST = [tnot(win(b,[m(a,b),m(b,c),m(c,d),m(d,a)]))] ? s

Stable Models: 
  {win(b,[m(a,b),m(b,c),m(c,d),m(d,a)]); win(d,[m(a,b),m(b,c),m(c,d),m(d,a)])} ? ;

  {win(a,[m(a,b),m(b,c),m(c,d),m(d,a)]); win(c,[m(a,b),m(b,c),m(c,d),m(d,a)])} ? ;
  no

no
nmr| ?-   
\end{boxedverbatim}
  \caption{Different queries against fixed program}
  \label{fig:win-notwin-rt}
\end{figure}

In the cases where the user knows enough about the non-relevant parts
of the program to know that they do not interfere in the stable models
of the relevant part, she can use XNMR as a fast Stable Models
computation system. Otherwise, XNMR computes supersets of the stable
models of the original program, restricted to the relevant part with
respect to the query given.  Alternatively, XNMR can be viewed as a
program generator for SMODELS. The program in
Figure~\ref{fig:win-notwin} is an example of such application. The
code dynamically builds programs according to the size and shape of
the graph passed to it as an argument. This program is then passed to
SMODELS for further evaluation.

\subsection{Application Program Interface}
\label{sec:api}

The top-level shell described above is implemented as an application
of the API provided by the XNMR architecture. The API provides
predicates to initialize the XSB/SMODELS interface, automatically
extracting the residual program for a given predicate, encoding and
sending it to SMODELS. Besides, several control and state-checking
predicates are defined. The most significant predicates provided by
XNMR's API are shown below.

\begin{itemize}
\item \texttt{init\_smodels(+Query)} --- initializes SMODELS with the
  residual program for Query; the residual program is collected from
  the table for Query, grounded if necessary, encoded and sent to
  SMODELS for further computation.
\item \texttt{atom\_handle(?Atom, ?AtomHandle)} --- is set by
  \texttt{init\_smodels/1} to be true of the set of atoms in the
  residual program (and thus the Herbrand Base of the Stable Models to
  be computed). AtomHandle is an integer uniquely identifying the
  atom. It is also used when decoding results returned from SMODELS.
\item \texttt{a\_stable\_model} --- invokes SMODELS to find a stable
  model of the residual program set by the previous invocation of
  \texttt{init\_smodels/1}. Fails if there are no (more) Stable
  Models. It will compute all stable models through
  backtracking. Atoms true in a stable model can be obtained by the
  next predicate.
\item \texttt{in\_current\_stable\_model(?AtomHandle)} --- true of
  handles of atoms that are true in the current stable model (set by
  an invocation of \texttt{a\_stable\_model/0}). 
\item \texttt{current\_stable\_model(-AtomList)} --- constructs a list
  of all atoms that are true in the current stable model (set by
  \texttt{a\_stable\_model/0}). 
\end{itemize}

\section{Conclusions}
\label{sec:conclusions}

In this work we introduce XNMR, a system that integrates two
computation engines for two different semantics of logic
programs. These engines are XSB, that computes the Well Founded
semantics of logic programs, and SMODELS, that computes Stable Model
semantics of a class of syntactically-restricted logic programs. 

The XNMR system provides an interactive environment for the
exploration of different semantics of logic programs. The interactive
nature invites uses such as debugging of knowledge bases and programs,
and the possibility to work with partial information provided by the
Well Founded semantics allows for the management of large systems,
exploiting the modularity of such systems.

The combination of the goal-directedness of XSB with the \emph{global}
nature of Stable Model semantics introduces some caveats with respect
to the semantics computed by XNMR. There exists a class of programs,
called call-consistent programs, for with XNMR computes only models
that have a one-to-one correspondence to the Stable Models of the
program. Still, we show that, even for programs not in this class, the
models computed by XNMR correspond to a class of Partial Stable Models
of the whole program, and can be useful in several applications. A
major advantage is the greater flexibility provided by the
unrestricted variables provided XSB. This allows, for instance, the
interactive exploration of predicates with infinite models, and the
creation of \emph{meta-programs}, that created ground programs to be
evaluated under the Stable Models semantics.

XNMR is distributed as part of the XSB System in the form of a package
that can be compiled when SMODELS is also present in the system. XNMR
can be obtained, along with XSB, from
\texttt{http://xsb.sourceforge.net/}.

%\bibliographystyle{plain}
%\bibliography{mybib}

\end{document}